\documentclass[a4paper,11pt]{article}
\usepackage{jinstpub} 
\usepackage{lineno,float,todonotes}

\title{\boldmath Strategy for Bayesian optimised Beam Steering at TRIUMF-ISAC's MEBT and HEBT Beamlines}

\author[a,b]{O. Hassan\orcidauthor{0000-0001-9021-5512},}

\author[a]{O. Shelbaya\orcidauthor{0000-0003-1796-3965},}

\author[a]{W. Fedorko\orcidauthor{0000-0002-5138-3473},}

\author[a,b]{T. Planche\orcidauthor{0000-0001-9657-4401},}

\author[a,b]{and O. Kester\orcidauthor{0000-0002-1809-5031}}

\affiliation[a]{TRIUMF, \\4004 Wesbrook Mall, Vancouver BC, V6T 2A3, Canada}
\affiliation[b]{Department of Physics and Astronomy, University of Victoria, \\Victoria BC, V8W 2Y2, Canada}

\emailAdd{oshelb@triumf.ca}

\abstract{In preparation for operation of multiple Rare Isotope Beams (RIBs) when the Advanced Rare Isotope Laboratory (ARIEL) becomes operational, TRIUMF embarked on a program of advanced beam tuning applications and machine learning tools. The strategy for operationalizing Bayesian optimisation for beam steering purposes is being developed. A previously reported centroid correction algorithm is used to tune accelerated charged particle beams at TRIUMF's ISAC postaccelerator facility. We present findings and results from multiple machine development experiments conducted between October and November 2024, as part of a pivot toward semi-automated machine tuning methods. These findings were instrumental in shaping the tuning strategy for the medium and high energy beam transport (MEBT, HEBT) lines at ISAC, by sequentially optimising sub-sections of the beamlines.}


\arxivnumber{2505.08767} 

\begin{document}
\maketitle
\flushbottom

\section{Introduction}

The Isotope Separator and Accelerator (ISAC) facility~\cite{Ball_2016} is built around a post-accelerator chain: An initial step Radio-Frequency Quadrupole (RFQ)~\cite{749942} accelerator linked to a Drift Tube Linac (DTL)~\cite{988305} by a medium energy section with a 90$^\circ$ corner, enabling a mass-to-charge ($A/q$) selection. The RFQ accelerates beam from 2.04 keV/u to an output energy of 153 keV/u, which is then injected into the DTL. The ISAC facility also receives stable beam through the Off-Line Ion Source (OLIS), which is a multi-configuration terminal comprising three ion sources \cite{Jayamanna2014}. The separated function \cite{laxdal1997separated} DTL allows for a fully variable energy output, starting at 153 keV/u to a maximum of 1.8 MeV/u. These linac properties divide the ISAC facility into three energy regimes:

    \begin{itemize}
        \item Low Energy Beam Transport (LEBT): 2.04 keV/u
        \item Medium Energy Beam Transport (MEBT): 153 keV/u
        \item High Energy Beam Transport (HEBT): 0.153-1.8 MeV/u
    \end{itemize}

In the past, tuning the ISAC beamlines was entirely done by manual adjustment of quadrupoles and corrective steerers, with a singular focus on the beam current transmission. This was a time-consuming procedure which led to highly variable tuning times to experiments that are not based on the design optics of the different linac sections. Due to the limited availability of operators and the simultaneous delivery of rare and stable isotope beams, this is not a feasible operation scheme for the ARIEL operation era at TRIUMF. Furthermore, higher availability of rare isotope beams is expected once TRIUMF’s ARIEL facility comes online \cite{ariel}. 

Consequently, a philosophical change in the operational tuning methodology used for beam delivery at ISAC is required, with a switch to more automated and model-based tuning techniques. Significant effort has been placed on simplifying the process of beam tuning and delivery to experiments, using a mixture of parallel modeling \cite{PhysRevAccelBeams.22.114602,PhysRevAccelBeams.24.124602} and optimisation software \cite{shelbaya2024tuning,rsi}. By utilizing knowledge of the beam dynamics design at ISAC and a speedy beam envelope code (\texttt{TRANSOPTR}~\cite{TRI-BN-16-06}), a complement of control room applications have been developed to augment the operators' capabilities and minimize tuning times to experiments. This will allow the facility to run more efficiently and help prepare the operators for the arrival of ARIEL.

\begin{figure}[!htpb]
    \centering\includegraphics[width=\textwidth]{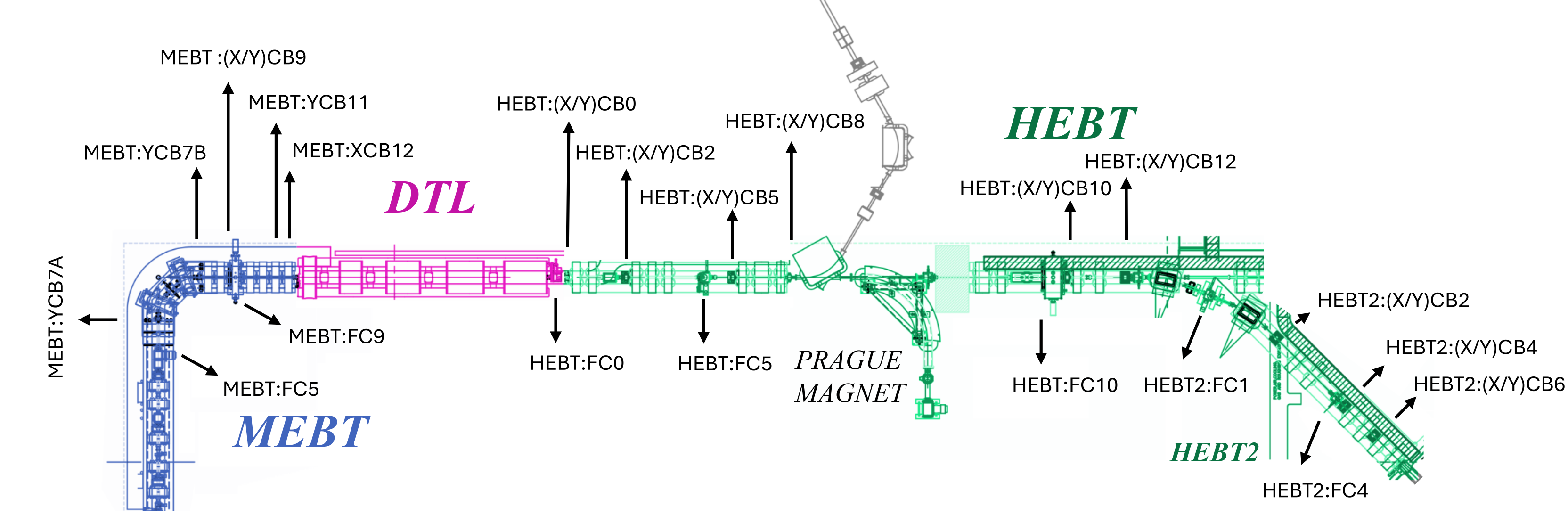}
    \caption{\label{fig:mebt_hebt} The MEBT to HEBT2 sections at the ISAC rare isotope beam facility, with the relevant steerers and Faraday cups highlighted.}
    \end{figure}

Rare isotope beam experiments can last anywhere from a few hours to several weeks. Since the required beam energy, mass-to-charge ratio, and species continually change between experiments, beam tuning at ISAC is carried out using a start-to-end model of the machine to calculate the necessary tune for achieving the desired beam parameters. This process is managed operationally through the Model Coupled Accelerator Tuning (MCAT) control room application~\cite{mcat}. However, obtaining the initial beam conditions at various operational sources (such as initial centroid position and divergence) can be challenging, which necessitates adjustments using corrective steering elements (see Figure \ref{fig:mebt_hebt}).

    
This report outlines significant upgrades to the operational tuning strategy for rare isotope beam delivery at TRIUMF's RIB postaccelerator. Over a two-month testing period, four distinct machine development tests were conducted, using several different beam species, with a focus on the medium and high-energy sections. The report discusses the optimal tuning strategies identified from the MEBT corner to the HEBT2 sections, including results from the machine development experiments.  The optimal steerer setpoints, which ensure improved beam transmission through the linac and beamlines, are determined through Bayesian optimisation~\cite{rsi}. It also presents the time required for tuning, along with recommendations for faster optimisation. Finally, the report outlines plans for future work to address both pre-existing and newly identified issues.

\section{Bayesian optimisation for Ion Steering}

\subsection{Bayesian optimisation}
Owing to diagnostic availability, finding the steerer values that maximize transmission is treated as a black box problem where each evaluation is costly. Bayesian optimisation is well suited for this problem, where the goal is to build a probabilistic model which accurately represents the objective function. This can be done by defining a surrogate model based on prior information, typically a Gaussian process (GP) \cite{rasmussen}. With a zero mean prior, a Gaussian process becomes entirely defined by the kernel. The kernel calculates the similarity between pairs of input data points to find the covariance matrix, which allows for the calculation of the posterior distribution. Bayesian optimisation for Ion Steering (BOIS) was developed at TRIUMF to optimise the corrective steerer problem at ISAC, with majority testing in the low energy beam transport (LEBT) section \cite{rsi}. The kernel used in BOIS is the Matérn Kernel \cite{matern1960kernel}. BOIS has an initial sampling stage where the inputs are randomly selected based on their mid-points, followed by the training stage where the optimisation is carried out.

\subsection{Optimising the Acquisition Function}
The acquisition function (AF) determines the selection of the next best input values to test. Optimising the AF is a critical step for the Bayesian optimisation approach. Typically there is a hyperparameter which controls the balance of exploration vs exploitation. Focusing on exploration biases the AF to find areas in the input space with high uncertainties, whereas exploitation focuses on areas presently known to produce high objective function value (and thus potentially missing a global optimum). A good balance is required to avoid getting stuck in a local maxima while ensuring efficient sampling of the input space.

\section{Tuning Sequences}

    

Since Bayesian optimisation is typically most effective for problems with smaller dimensions~\cite{frazier2018tutorialbayesianoptimization}, beamline segments for optimisation must be carefully selected to keep the number of optimisation variables manageable, typically under 20. By dividing the full machine optimisation into several smaller segments, each overlapping with the previous one, greater efficiency and reliable operation are achieved. These segments are selected based on the availability of diagnostics and the typical operator tuning procedures. Each sequence  includes steering elements and beam current monitors, such as Faraday Cups (FC), although some sequences, such as the MEBT section, also incorporate quadrupoles to account for modeling uncertainties, which are being separately investigated. Further discussion of this necessity is presented in~\cite{TRI-BN-22-29,Shelbaya:2023jcm}, however the determined bounds for BOIS's quadrupole adjustments are $\pm10\%$ of the MCAT computed quadrupole gradients. 

The first sequence starts in the MEBT corner and goes through the DTL to the start of HEBT (Fig.~\ref{fig:mebt_hebt}, MEBT:FC5 to HEBT:FC5). This is the only sequence where quadrupoles are tuned due to disagreements between the model and the observed tune. None of the DTL or HEBT elements are tuned in this sequence, it is purely a DTL injection optimisation problem. Sequences 2-5 cover HEBT and HEBT2, sequence 2 is found to be trivial for the optimiser and will likely be combined with sequence 3 in the future.  


\section{Results}    

    \subsection{MEBT Corner and DTL}



\begin{figure}[!htpb]
    \centering\includegraphics[width=\textwidth]{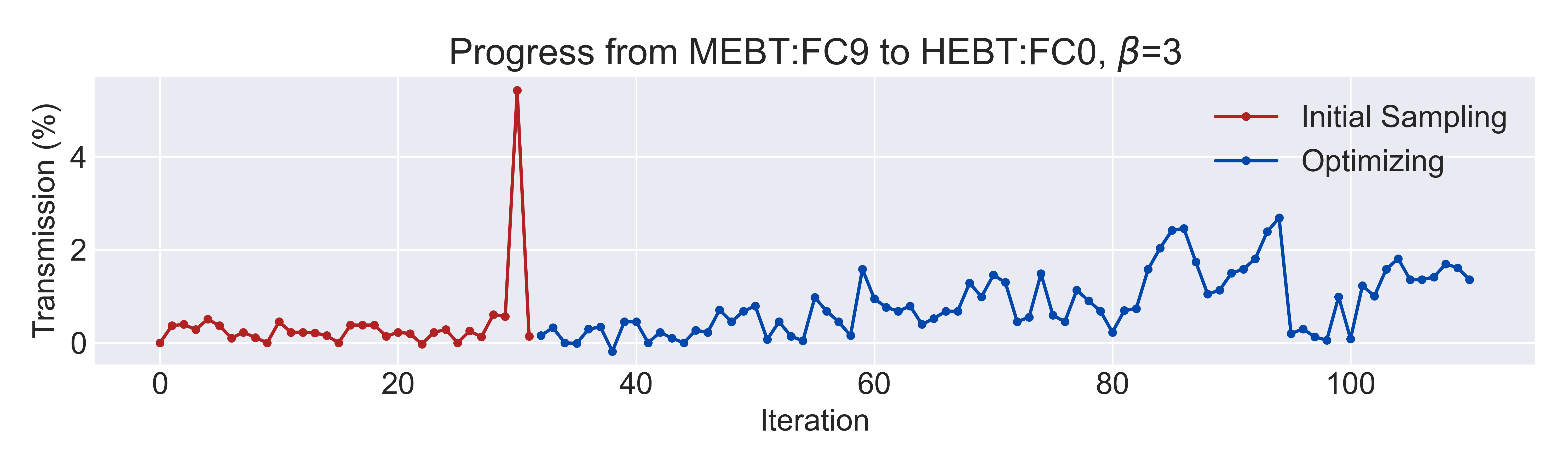}
    \centering\includegraphics[width=\textwidth]{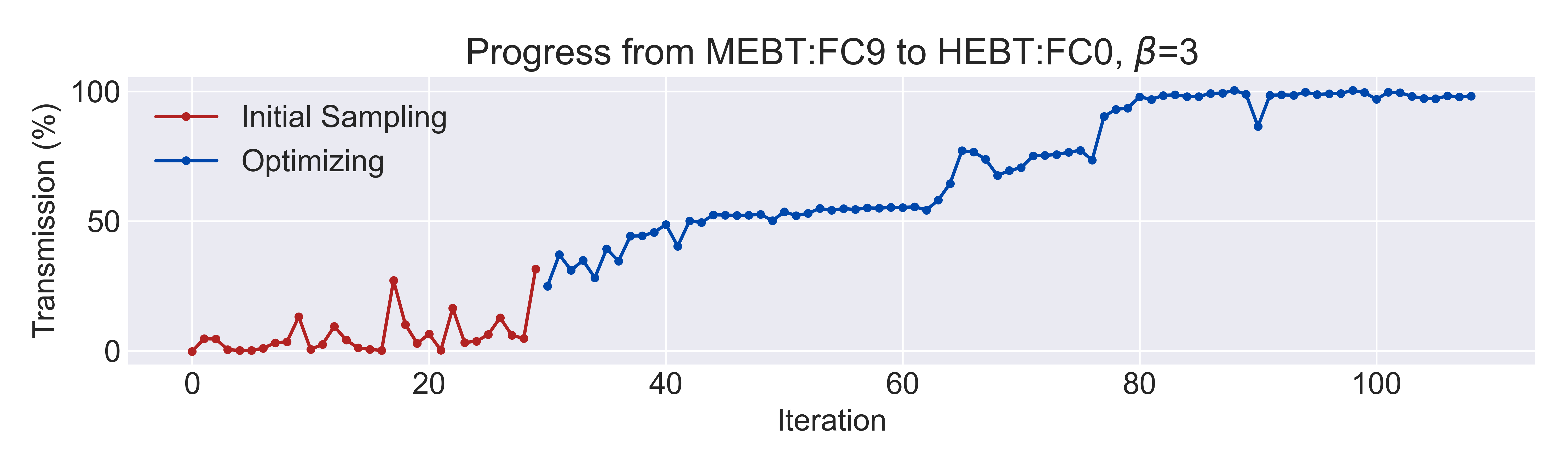}
    \caption{\label{fig:mebt} \textbf{Top:} optimisation with quadrupoles unbounded. \textbf{Bottom:} Bounding quadrupoles to $\pm10\%$ of the MCAT value.}
    \end{figure}

    
    The beam species was $^{84}$Kr$^{15+}$ from the multicharge ion source (MCIS) terminal at OLIS, accelerated by the DTL to E/A=461 keV/u. This development period was focused on testing different sequence definitions and optimising through the MEBT corner. The objective was to maximize beam transmission as measured from one Faraday cup to the next. The optimisations were performed using the Upper Confidence Bound (UCB)~\cite{srivanas2010ucb} acquisition function where the exploration vs exploitation hyperparameter ($\beta$\footnote{In this work, $\beta$ refers exclusively to this hyperparameter.}) was kept at $\beta=3$, which was found to provide highly robust performance. The initial tests excluded all quadrupoles, which failed to exceed 65\% transmission across the DTL. Figure \ref{fig:mebt} shows the relevant tests where quadrupoles were included in the optimisations. Keeping them unbounded was unsuccessful, producing 5\% transmission across DTL. Bounding the quadrupoles to $\pm10\%$ of the MCAT values yielded a DTL transmission of 100\%. Finally, note:
    \begin{enumerate}
        \item Operators noticed one of the MEBT corrective steerers tripped repeatedly during the tests, which could be related to unipolar power-supply and polarity switch operation \cite{TRI-BN-24-12}. 
        \item Current entering the MEBT section fell from 6.2nA at the start of testing to 5.5nA by the end, likely due to ion source output current variation.
    \end{enumerate}

    \subsection{HEBT to HEBT2}
    


Tests were carried out using both Expected Improvement (EI)~\cite{ei} and UCB, it was found that EI generally converges faster but is less consistent.  Transmission optimisation was achieved using sequences 2-5. The Bayesian optimiser was run with EI and $\beta$=10, which produced high transmission for Sequence 2. However, note that this is a straight beamline segment featuring a relatively large aperture (2.54\,cm beampipe radius) and an absence of notable constrictions. High transmission is found initially during the random sampling stage, before an acquisition function is even used. 

\begin{figure}[!htpb]
    \centering\includegraphics[width=\textwidth]{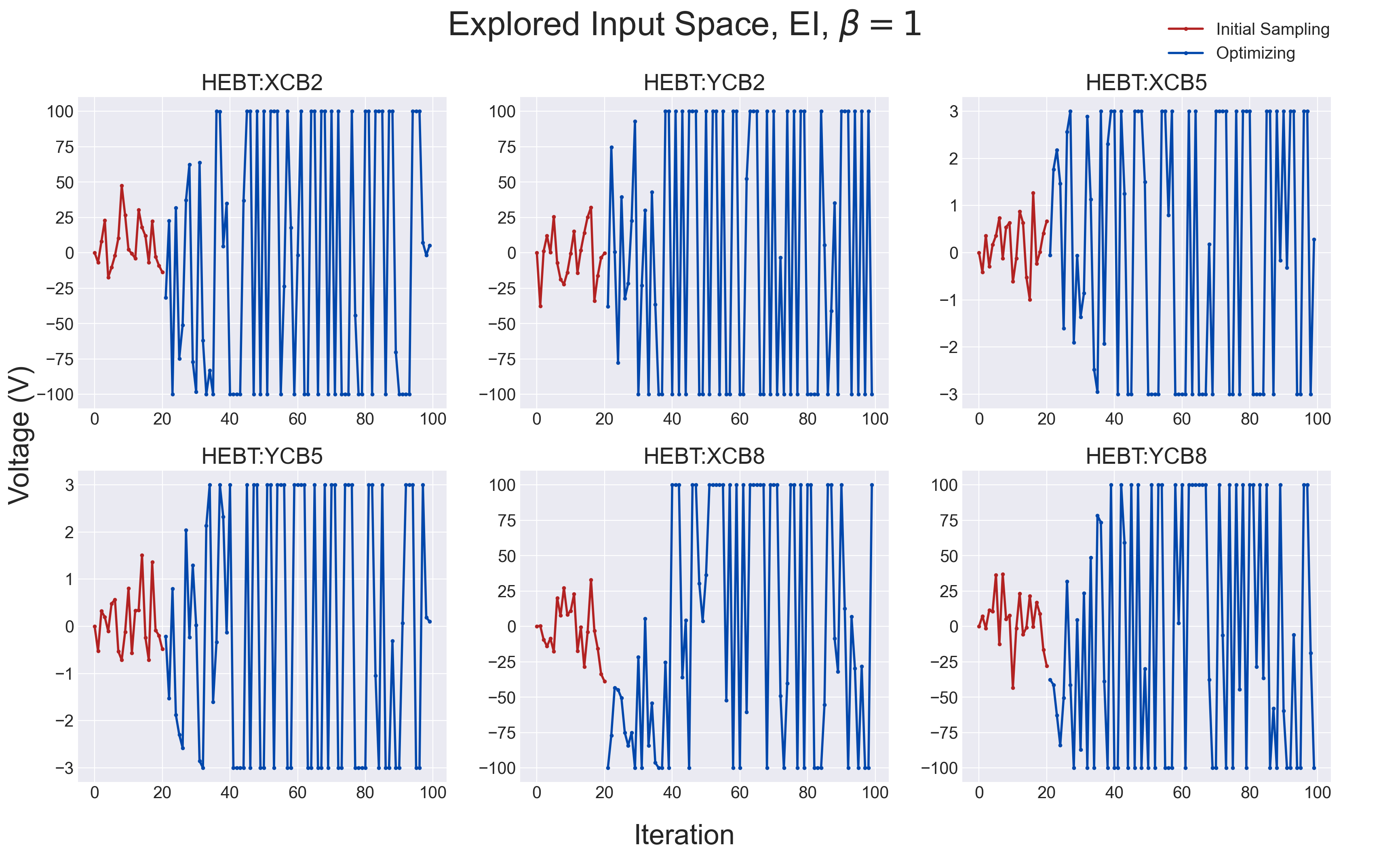}
    \caption{\label{fig:overexplore} Explored input space during optimisation for EI with $\beta=1$.}
    \end{figure}

Sequence 3 was more challenging due to the presence of bunching cavities, which remained unpowered, whose drift tubes possess a 1\,cm aperture radius \cite{TRI-BN-19-06}, compared to the 2.54\,cm beampipe radius. Figure \ref{fig:overexplore} shows the explored input space with expected improvement for $\beta = 1$. 
    
\begin{figure}[!htpb]
\centering\includegraphics[width=\textwidth]{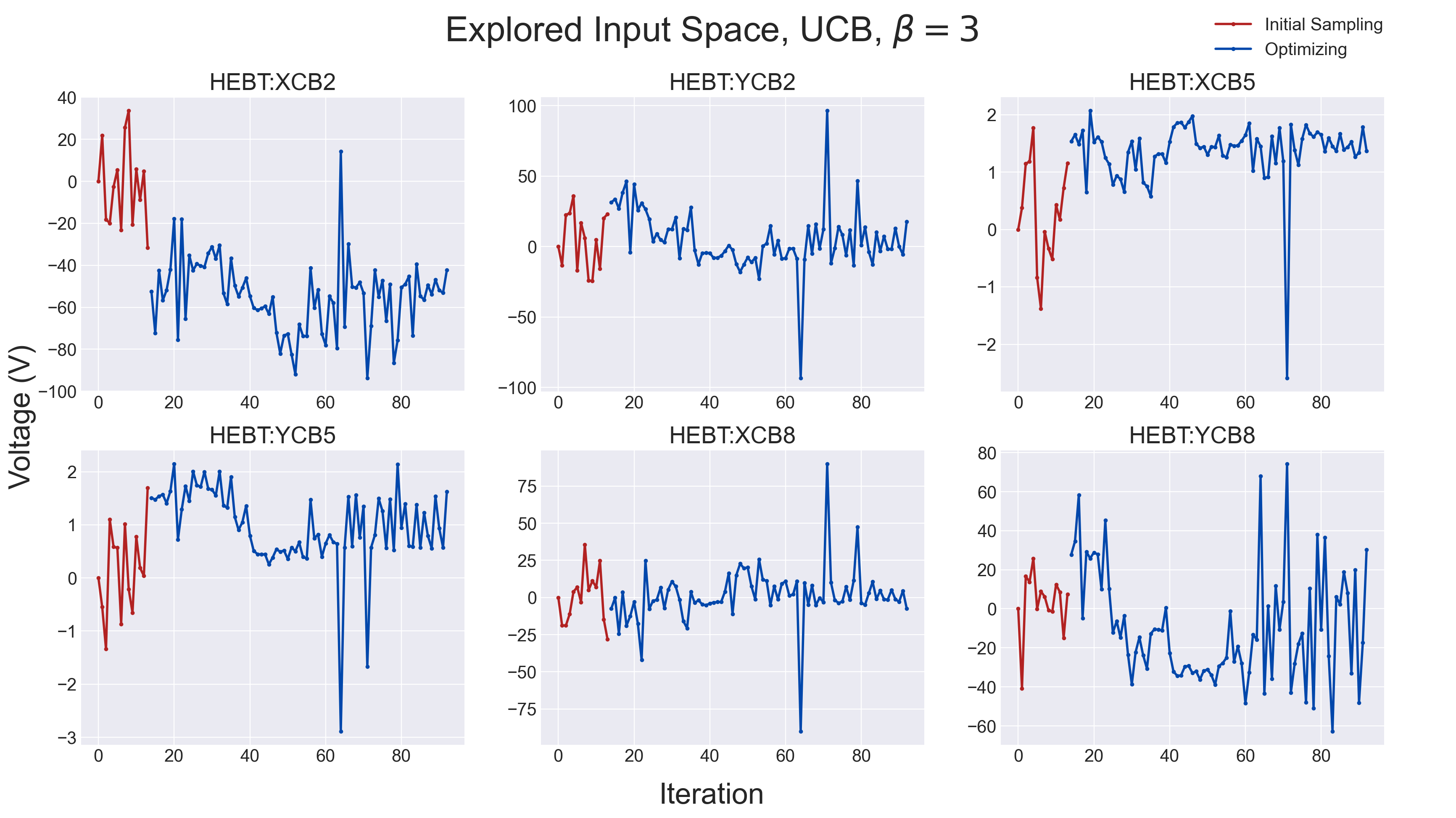}
\caption{\label{fig:overexplore2} Explored input space during optimisation for UCB with $\beta=3$.}
\end{figure}

In this case, the BOIS algorithm failed with significant over-exploration. As a result, transmission was at a minimum, in the $\approx 1-5\%$ range. This issue was later found to be with the \texttt{BoTorch} package \cite{balandat2020botorchframeworkefficientmontecarlo}: The previously employed version was {\tt 0.11.3}, however upgrading to version {\tt 0.13} fixed this issue. A future investigation will look at these acquisition functions and the issues encountered in more detail, however it is noted that strict version adherence will be necessary to avoid functionality problems. As a result, it was resolved to employ UCB for this section with $\beta=3$. This configuration proved the most reliable, delivering a measured transmission of 93\% through this section. Figure \ref{fig:overexplore2} shows the expected input space exploration.

\newpage

\begin{figure}[!htpb]
\centering\includegraphics[width=0.89\textwidth]{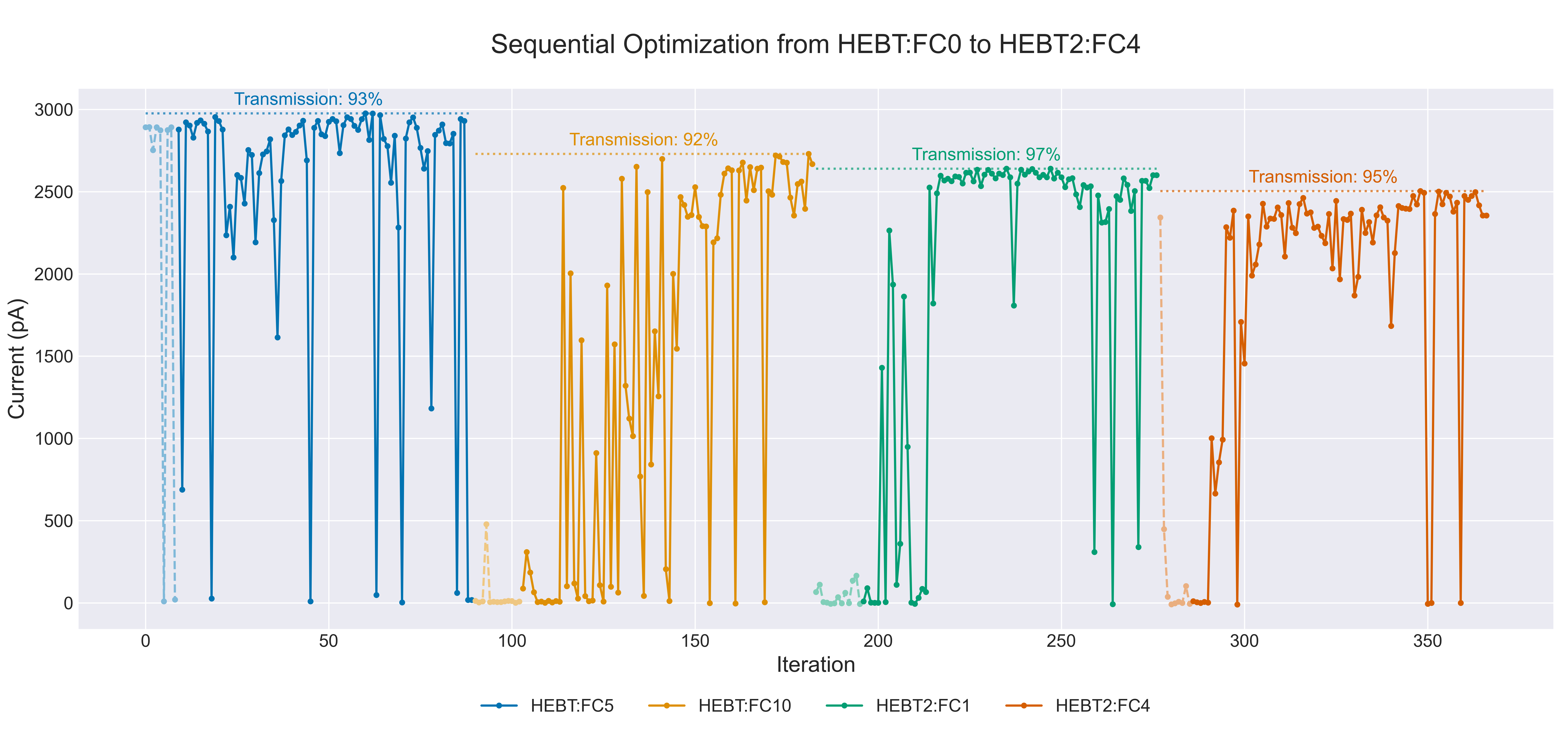}
\caption{\label{fig:sequential} 4 separate optimisation problems from HEBT:FC0 to HEBT2:FC4. Sequences 2-5 are covered with the best transmission shown for each sequence.}
\end{figure}

Sequences 4-5 were optimised with UCB, shown in Figure \ref{fig:sequential}. Before early-stopping was introduced, the total tuning time lasted $\approx42$ minutes. This can be reduced with future development, as the maximum transmission is often found very early during the optimisation. Implementing an early stoppage option will address this. A full implementation of early stoppage requires two criteria: 1) Beam transmission reaches a pre-determined threshold. 2) No improvement in beam transmission after a given number of iterations. It is also possible to set the hyperparameter $\beta=0$ to maximally exploit the best inputs for $\sim$5 iterations after a condition is met.

\newpage

    \subsection{Combining Sequences 2-5}


    \begin{figure}[!htpb]
    \centering\includegraphics[width=\textwidth]{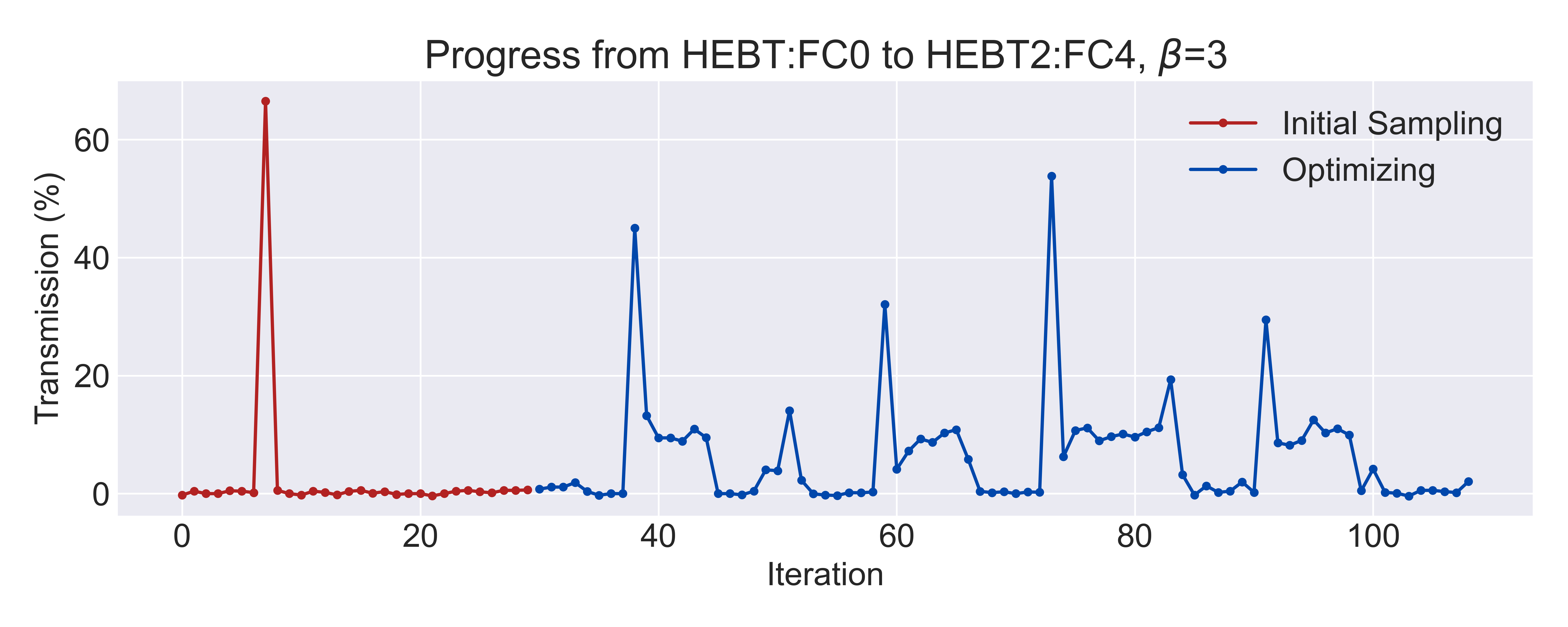}
    \caption{\label{fig:hebt_full} Single optimisation problem where Sequences 2-5 are combined.}
    \end{figure}

    To demonstrate the inefficacy of optimising long stretches of the beamline with a large optimisation problem, a test was conducted where Sequences 2-5 were combined into a single optimisation problem, involving 14 corrective steerers. Figure \ref{fig:hebt_full} illustrates the results, with the most noticeable observation being the lack of improvement over iterations. The best point reached during random sampling was never exceeded, and the final transmission from HEBT:FC0 to HEBT2:FC4 was 68\%, significantly lower than the 86\% achieved using smaller sequences. 
    

    \subsection{Tuning Time Estimates}



     With a beam species of $^{12}$C$^{3+}$ (Table 2, Test 1) at 464 keV/u, another test was carried out with the goal of recording a time-estimate of BOIS transmission optimisations from Sequences 1-5 using early-stopping. Manual stoppage was used when transmission reached 90-95\% or ceased to improve. Table \ref{tab:tuning_times} shows the tuning times per sequence. Sequence 3 was again more challenging for BOIS to optimise. Matching through the HEBT bunchers is a likely cause for the transmission difficulties due to the aperture restrictions of the HEBT bunchers \cite{TRI-BN-19-06}. Further scrutiny of this section will be carried out in the future to better understand and address this.

     \begin{table}[h]
    \centering
    \caption{Tuning Times per Sequence (elapsed time between start and finish)}
    \label{tab:tuning_times}
    \begin{tabular}{c c c}
    \hline
    \textbf{Sequence No.} & \textbf{\# of Elements} &  \textbf{Tuning Duration (min)} \\
    \hline
    \#1 & 14 & 9 \\
    \#2 & 4  &  1 \\
    \#3 & 6  &  6 \\
    \#4 & 6  &  5 \\
    \#5 & 4  &  2 \\
    \hline
    \textbf{Total} &  &  28 \\
    \hline
    \end{tabular}
    \end{table}

    \subsection{Further Tests}

    Two final tests were carried out, first using $^{12}$C$^{3+}$ (Table 2, Test 2) at 464 keV/u, and then $^{133}$Cs$^{21+}$ at 254 keV/u, both from the MCIS source. The carbon beam test suffered from current degradation due to an ion source failure, leading to a reduction in available beam for testing. The caesium beam also suffered from similar source issues. In both cases, significant ($>$20\%) beam current fluctuations were recorded at irregular intervals. Testing proceeded during periods of relative beam stability. These instabilities originate in the ion source and were discovered when using BOIS. As this is outside of BOIS's control, this issue is being addressed through further studies of the ion source.




    
    \subsection{Summary}

    BOIS consistently achieves high transmission rates across the medium and high energy sections while maintaining a short tuning time, provided the problem is suitably divided into appropriate sub-sections for optimisation. This has been tested with multiple beam compositions, mass to charge ratios and final energies. These results, listed in Table \ref{tab:beam_efficiency_transposed}, will inform the development of BOIS as a control room application for operator use when tuning the ISAC post-accelerator for beam delivery. With the aforementioned methodology, tuning from the MEBT corner to HEBT2 takes 28 minutes, which is comparable to the tuning times achieved by the operators. This technique we presented is generalisable at TRIUMF in several areas. These are the H$^-$ injection line for the 520\,MeV cyclotron, the ISAC-2 superconducting (SC) linac, and the new ARIEL beam lines including the SC electron linac. This works also for RIB facilities as the only diagnostic requirement is a Faraday cup to provide continuous readback of the beam current.

    \begin{table}[!htpb]
      \caption{Current transmission attained by BOIS across different sequences for various beam species and final energies. Note that the sequence data was collected over several independent runs rather than in a single, strictly ordered sweep.}
      \centering
      \small
      \resizebox{\textwidth}{!}{%
        \begin{tabular}{l c c c c c c}
          \hline
          \textbf{Species} & \textbf{Energy (keV/u)} & \textbf{Section 1} & \textbf{Section 2} & \textbf{Section 3} & \textbf{Section 4} & \textbf{Section 5} \\
          \hline 
          $^{84}$Kr$^{15+}$         & 461 & 100\% & 93\%  & 91\%  & 97\%  & 95\%  \\
          $^{12}$C$^{3+}$ [Test 1]    & 464 & 89\%  & 100\% & 90\%  & 94\%  & 97\%  \\
          $^{12}$C$^{3+}$ [Test 2]    & 464 & 76\%  & 98\%  & 78\%  & 82\%  & 92\%  \\
          $^{133}$Cs$^{21+}$         & 254 & 100\% & 100\% & 97\%  & 91\%  & 97\%  \\
          \hline
        \end{tabular}%
      }
      \label{tab:beam_efficiency_transposed}
    \end{table}
    
\section{Future Work}
    \label{future}

    While these results are very promising, there are several improvements required for fully reliable operation, with some requiring machine development shifts for testing:

    \begin{itemize}     
        \item Implement early stoppage as a part of BOIS with adaptive $\beta$, this will be especially useful once the control room application is deployed.

        \item Minimize steering by bounding the inputs, this has been developed with boundBOIS, reported in~\cite{rsi}, but not extensively tested in the medium to high energy beamlines.

        \item Include the best input values from previous runs in the initial sampling stage, which are scaled for beam $A/q$ for the present species. Currently, initial sampling uses a random process which calculates test values from the midpoint of the bounds.
        


        \item The beam current as measured on the Faraday cup is averaged over 20 measurements, this is redundant as the wait time between each measurement (0.01s) is far quicker than the response time for the control room application webserver (0.2s). It is best to average over a smaller number of measurements and provide the proper wait times between the measurements.

        \item The Faraday cup noise is currently treated as a learned hyperparameter during the optimisation process. A more accurate method could be to explicitly calculate the variance of the Faraday cup measurements and pass it along to the Gaussian process.

    \end{itemize}

\acknowledgments

Thanks to Chris Charles for providing beam time after the ASPIRE~\cite{aspire} experiment finished early. Credit to Rick Baartman for helping determine the procedure and assisting with analysis. Thanks to all the RIB operators for their support during development. An LLM was used to refine the writing in certain sections. TRIUMF is located on the traditional, ancestral, and unceded territory of the Musqueam People, who for millennia have passed on their culture, history, and traditions from one generation to the next on this site. We acknowledge the support of the Natural Sciences and Engineering Research Council of Canada (NSERC), [funding reference number SAPPJ-2023-00038]. 




\bibliographystyle{JHEP}
\bibliography{biblio.bib}






\end{document}